\documentclass[a4paper]{VisionStyle}
\usepackage{epsfig}

\begin{document}

\title{Active Galaxies in the XMM/Chandra Era}

\author{R.F.\,Mushotzky} 

\institute{
NASA/Goddard Space Flight Center, Code 662, Greenbelt, MD 20771, USA}

\maketitle 

\begin{abstract}

In this review talk I discuss the important issues of AGN research and how
the new generation of X-ray observatories can help to constrain the physics
of AGN.  I also present a biased list of the new {\it XMM-Newton} and 
{\it Chandra} discoveries and how they
have altered our view of AGN. I also present a set of what are the new types
of observations that need to be performed to make significant progress in
this field.

\end{abstract}

\section{Introduction}

Why are active galaxies interesting in the X-ray?
Active galaxies (often abbreviated as  AGN, active galactic nuclei) are the
most numerous class of extragalactic X-ray source (above a flux limit of
$F(x)>10^{-14}$ ergs cm$^{-2}$ s$^{-1}$). With the extremely 
sensitive CCD cameras of {\it XMM-Newton}
and {\it Chandra} they  have been  detected  out to $z\sim6$ 
(\cite{rmushotzky-C2:vea01}), and
potentially further. Their intrinsic luminosity covers an extremely wide
range from $<10^{40}$~ergs~s$^{-1}$ (\cite{rmushotzky-C2:hoea01}) to over 
$10^{47}$~ergs~s$^{-1}$ (\cite{rmushotzky-C2:fea97}) 
with the ``X-ray'' (0.1-100 keV) band having ~0.05-0.3 (all?) of the
total energy.  The X-rays originate very close to the supermassive black
hole (MBH) as shown by the fact that the  X-ray band  is the most ``rapidly''
variable of all wavelength bands. The implication is that rapid variability
indicates a small intrinsic size and that a small size is associated with
the smallest of all emission regions, the black hole itself.  

\begin{figure*}[ht]
  \begin{center}
    \epsfig{file=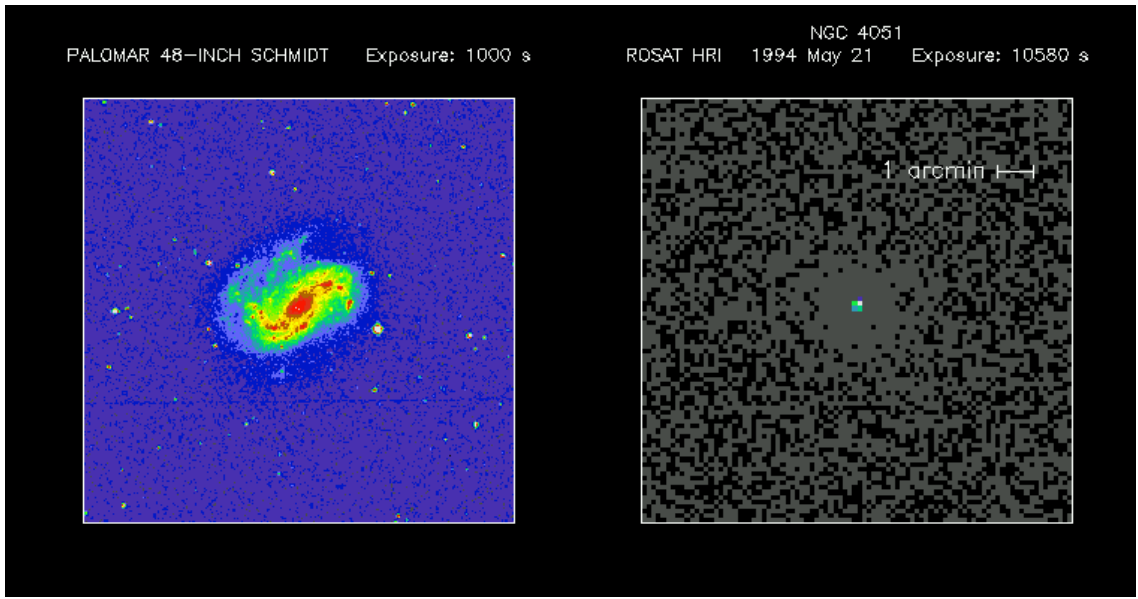, width=15.0cm}
  \end{center}
\caption{Left Panel: a Digital Sky Survey image of the low 
luminosity nearby Seyfert I galaxy NGC~4051. Right Panel: a 
{\it ROSAT} HRI observation of the same object.  Notice the 
enormous increase in constrast with the X-ray image and the ease
with which the central source is detected.}  
\label{rmushotzky-C2_fig:ngc4051}
\end{figure*}

The X-ray band  has the only spectral signature that  originates  close to
the MBH itself -- the broad ``Fe K'' line (e.g., \cite{rmushotzky-C2:nea97}). 
One of the surprises of the {\it XMM-Newton} and {\it Chandra} mission 
is that the X-ray band is the most efficient way of finding AGN 
(Figure\ref{rmushotzky-C2_fig:ngc4051}). This is shown by the
ease with which X-ray emission is detected in objects that have ``normal''
optical, UV, IR and radio properties and the very large number of objects
per unit solid angle detected by {\it Chandra} and 
{\it XMM-Newton} (more than 1000 deg$^{-2}$, 
\cite{rmushotzky-C2:muea00}; \cite{rmushotzky-C2:hea01}) which are almost
certainly active galaxies compared to optical surveys which find fewer than
100 deg$^{-2}$ (\cite{rmushotzky-C2:kea97})

In addition to ``direct'' emission from the region of the MBH itself there
is significant X-ray radiation from ``jets'', both in the plane of sky as seen in the
extended X-ray emission from jets and from jets in the line of sight
(theses objects are called ``BLAZARS'') which is evident in the very high
luminosities of these gamma-ray bright objects. All types of AGN are  luminous
X-ray sources:  both radio quiet and radio loud, those with broad optical
lines and those with narrow lines. In fact  the whole veritable zoo of
active galaxy names  (Bl Lac, Quasar, Seyfert I/II, LINER, NLAGN, BLRG)
are luminous X-ray sources. In addition  there is X-ray emission from objects which
show no indication of activity in any other wavelength band; that is
many/most X-ray selected AGN show weak or no optical ``peculiarities''
(\cite{rmushotzky-C2:mea00}).

\section{What are Active Galaxies?}
\label{rmushotzky-WA2_sec:what}

It is almost certain that most active galaxies are powered by the energy
released due to accretion onto a MBH, however so far there has been no
direct observation of the process of accretion.  In addition there may well
be other  processes  at work such as the extraction of the spin energy of
the black hole (\cite{rmushotzky-C2:wea01}) or  the effects of magnetic fields.
While we believe that most of the energy comes from accretion the exact
mechanism which produces X-rays is not known. This lack of knowledge of the
physical mechanism or the geometrical relation of the MBH to the X-ray
source is one of the prime difficulties in understanding the wealth of
available data. 

In recent years there has been the accumulation of strong dynamical
evidence for MBH from the motions of stars, ionized and cold gas in  the
nuclear regions of many nearby galaxies (e.g., \cite{rmushotzky-C2:kg01}). 
It seems as if the mass estimates are accurate to $\sim2-4$ for a sample
of over 20 objects. What is peculiar for many of us who have been in the
field for years is that all but one of the  proven massive black holes are
in non-active  galaxies, while the whole paradigm for the existence of MBH
was driven by the existence of active galaxies! These recent studies have
shown that there are strong connections between the spheroidal component of
the host galaxy and the MBH, such that the mass of the MBH can be
accurately estimated from the velocity of the stars in the spheroid  and
roughly estimated by the luminosity of the spheroidal component of the
galaxy. There appears to be little relation between the stars or gas in the
disk and the mass of the MBH. This is rather interesting since most active
galaxies in the local volume of space are spiral systems which do not seem
to have a luminous bulge. Since most of the objects are not active at the
present time, the mass of the MBH must have been grown much earlier and be
strongly connected to how the galaxy forms. 	

The ratio of the photon luminosity to the mass of the object, the
Eddington ratio, ranges from $<10^{-7}$ to $>1$ with some of the lowest values
coming from {\it Chandra} observations of nearby giant elliptical galaxies
(\cite{rmushotzky-C2:loea01}). 
Recent estimates of the mass of black holes in active galaxies is
consistent with that in non-active galaxies (\cite{rmushotzky-C2:fea01})  and
the new mystery is why so many MBHs are so quiet (e.g. non-active). 

It is clear that relativistic effects might be dominant in the X-ray
emission and are very  important in radio loud AGN. This is shown by the
very rapid X-ray variability, the direct connection of radio, X-ray and
gamma-ray emission in BLAZARS and the {\it Chandra} observations of X-ray
emission from spatially resolved jets 
(\cite{rmushotzky-C2:saea01}) . This direct
emission from the jets, shows that , in some sources, a considerable
fraction of the total X-ray flux may not come from near the nucleus. 

\section{What are the Fundamental Questions?}
\label{rmushotzky-WA2_sec:fund}

After over 50 years of study of active galaxies and more than 25 years of
detailed study in the X-ray band many fundamental questions remain. 

1) How do AGN ``work'' -- e.g., how is energy produced/extracted and transformed
into radiation. While there has been considerable progress in understanding
the physics of accretion and how energy can be extracted from accreting
matter, there has been little progress in understanding the process of
X-ray photon production or how the energy released in accretion can be
transformed into X-ray radiation. It is generally believed that the X-rays
in luminous objects are produced by ``thermal Compton scattering'' 
(\cite{rmushotzky-C2:peea01}) where low energy photons are upscattered into the
X-ray band via scattering off of energetic electrons. A detailed analysis
of this process in galactic black holes shows that there is great
difficulty in understanding the details of the time and spectral
variability (\cite{rmushotzky-C2:kcg01}). Both the physical location of the
scatterer and the understanding of how the energy is partitioned into
different energy bands are uncertain. It is clear that the X-ray and UV/IR
bands have a very complex relationship since the optical/UV band does not
vary as rapidly as the X-ray band (\cite{rmushotzky-C2:nea98}).

2) How is the MBH  connected to host galaxy? That is  how do MBHs  form and
effect the galaxy in which they live. Because the mass of the black hole,
but not its luminosity is closely related to the stellar properties of the
galaxy, the growth and formation of the MBH must be closely related to the
formation of the stellar spheroid. Since  X-ray surveys find many more AGN
than optical surveys and are sensitive to objects embedded in large column
densities of material they may  be the best way to find MBHs in forming
galaxies at high redshift where other techniques cannot detect AGN because
of the effects of redshift and the Lyman-$\alpha$ forest.

3) What is the origin of the wide range of apparent types?
The  incredible range of radio-IR-optical-UV-X-ray and gamma-ray properties
of AGN have produced a zoo of names. While there have been many attempts at
unification of objects via geometrical effects (so-called unification
models, \cite{rmushotzky-C2:ant93}) 
with a fair degree of success it has become clear
that there is a very poor relation between the optical and X-ray properties
of objects and thus the whole concept of unification of optically
classified objects is called into question. There seems to be some
additional property needed: perhaps the effect of star forming regions
(\cite{rmushotzky-C2:lwh01}), intrinsic weakness of the optical continuum as seen
in some nearby LINERS such as M81 (\cite{rmushotzky-C2:hfs96}) or the effects of an
ionized absorber (\cite{rmushotzky-C2:bfp96}) 
such that high X-ray column densities
and low optical reddening and UV absorption are both present.

The apparently fundamental dichotomy between radio bright and radio quiet
objects seems to be related to the mass of the MBH, with more massive
galaxies, with presumably more massive MBHs having a higher probability of
hosting a radio loud AGN (\cite{rmushotzky-C2:dea01}).

4) How do they evolve with cosmic time? -- It is clear from the faint source
counts (\cite{rmushotzky-C2:mg01}) and the redshift distribution of the
X-ray selected sample (\cite{rmushotzky-C2:lfea01}) that there is strong
evolution in luminosity and numbers of X-ray selected active galaxies.
Since the number density of low redshift MBHs is much higher than that of
AGN this increase in objects and luminosity with cosmic time presumably
represents the epoch when these MBH were luminous. The origin of this
phenomenon, that is what is actually evolving,  and whether these objects
are fundamentally different from low  z objects is not yet clear.

5) What can we learn about strong gravity? Because the Fe K and perhaps
other X-ray spectral features, as well as the continuum, originates from
close to the central object there is potential for studying the effects of
strong gravity near the black hole (e.g., \cite{rmushotzky-C2:mea01}). There are
strong indications from the shape of the Fe K line in many (but not all)
Seyfert galaxies and galactic black hole candidates that the effects of
special and general relativity are important. It is not yet clear how these
features can be unfolded from the unknown geometry of the central regions
to learn about strong gravity, but the effects are clearly there. It is
also not clear why some objects do not show such Fe K lines, but studies of
galactic black hole candidates indicate that the structure of the accretion
disk may change drastically with little or no change in source intensity. 

6) What is the geometry of the central regions? While we have a general
cartoon idea of what the central regions of  active galaxies look like,
(e.g., \cite{rmushotzky-C2:muea93}) it is not a well tested paradigm.
The shape of, or even the existence of structures like  an accretion disk,
an obscuring torus or a coronae are not well determined. Without a better
understanding of the geometry of these objects much of our data is
simply uninterpretable. 

\section{{\it XMM-Newton} and {\it Chandra}}
\label{rmushotzky-WA2_sec:xmmchandra}

These missions represent a vast improvement in  a wide parameter space for
observations of AGN: \\
a) grasp -- that is high collecting area over a broad bandpass \\
b) high spectral resolution- with a factor of 100 improvement in energy
resolution over {\it ASCA} and a factor of 100 improvement in collecting area
over the previous high resolution spectrometers on {\it Einstein} \\
c) multiwavelength capability- the optical monitor on {\it XMM-Newton} allows
simultaneous measurement of the UV spectrum with a grism, or higher time
resolution continuum monitoring for a large number of active galaxies. \\
d) angular resolution (with the {\it Chandra} 
FWHM$\sim0.25''$) allowing the first
detailed X-ray images of the nuclear regions, precise positions and the
measurement of extended small scale emission. \\
e)  sensitivity allowing detection of sources $\sim100\times$ fainter 
than before, as well as an increase in the time resolution and spectral 
measurement flux threshold. \\

These properties make the {\it XMM-Newton}/{\it Chandra} 
combination  ideally suited for: \\
1) extending previous studies  to fainter, higher redshift, higher  and
lower luminosity systems \\
2) detailed studies of many bright low redshift objects \\
3) critical progress in the temporal/spectral domain \\
4) building up a complete picture of the multi-wavelength spectrum of
active galaxies as a function of redshift, type and luminosity, probing the
evolution of quasars  over the lifetime of the Universe. \\
5) examining the  structure of the central engine in  Seyfert galaxies via
observations of their broad band X-ray spectrum as well as the
multi-wavelength spectral properties and time dependent spectral signatures.

\section{Exciting New Observations (A Biased List)}
\label{rmushotzky-WA2_sec:newstuff}

1) The existence of narrow absorption lines that are the strongest features in
grating observations of  Seyfert I galaxies (\cite{rmushotzky-C2:kea00}; 
\cite{rmushotzky-C2:kea01}) while the 
total opacity is dominated by edges. While emission lines
are present they are weak. This is to be compared to the optical, IR and
UV spectra  which are dominated by emission lines. This radical difference
between the X-ray and other wavelength bands has not been fully understood
yet and indicates the absence of a X-ray ``broadline region''. This
fundamental difference maybe related to the existence of the classical
``two-phase'' instability such that the ionization parameter range where the
X-ray emitting gas might exist is unstable, while the optical and UV
emitting gas lies in a stable phase (\cite{rmushotzky-C2:kiea01}). 

\begin{figure}[ht]
  \begin{center}
    \epsfig{file=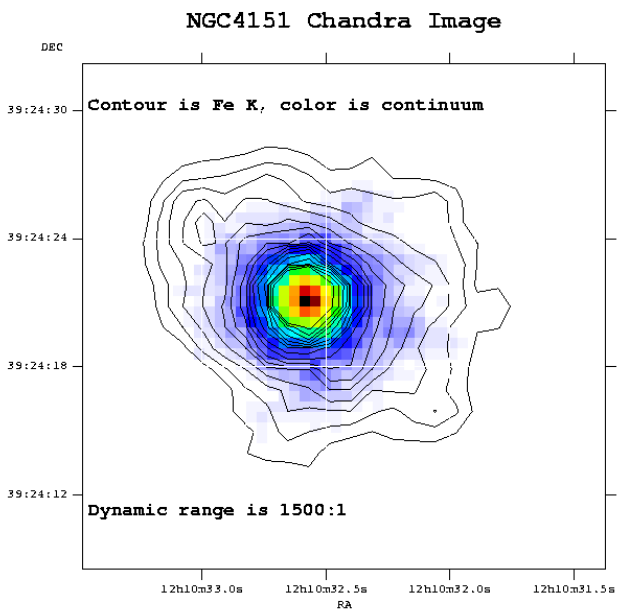,width=8.5cm}
  \end{center}
\caption{The {\it Chandra} high resolution image of the 
continuum in the 3-5 keV band (color) and the Fe K band (6.2-6.8 keV, 
contour) for the bright Seyfert I galaxy NGC~4151. Notice that both are 
extended but that the Fe K band seems to be more extended. The very high 
dynamic range (500:1) allowed by the {\it Chandra} sharp point spread 
function is necessary to detect the extended
emission.  The physical scale is $\sim500$ pc/per tic mark.}  
\label{rmushotzky-C2_fig:ngc4151}
\end{figure}

\begin{figure}[ht]
  \begin{center}
    \epsfig{file=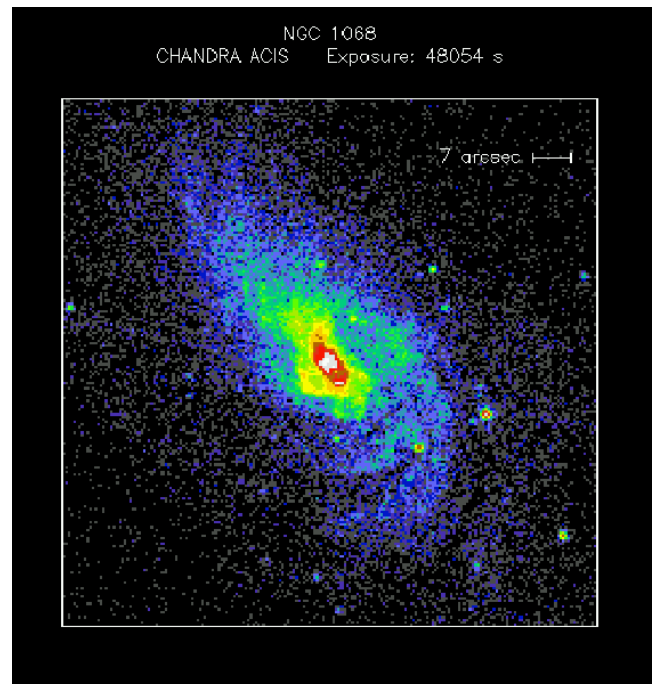,width=8.5cm}
  \end{center}
\caption{The {\it Chandra} image of the central $30''$ 
of the nearby Seyfert II galaxy NGC~1068.  Notice the very high 
surface brightness, extended emission from the central
regions and the diffuse extended galaxy wide emission. The bright sources
to the south-west are most likely ultra-luminous sources in NGC~1068.}  
\label{rmushotzky-C2_fig:ngc1068}
\end{figure}

2) The detection of extended line emission from O~{\small VII} 
and Fe K  in at least two AGN (\object{NGC~4151}, 
Figures~\ref{rmushotzky-C2_fig:ngc4151},\ref{rmushotzky-C2_fig:ngc1068}, 
\cite{rmushotzky-C2:oea00}; \object{Circinus}, \cite{rmushotzky-C2:sm01}) 
and extended soft X-ray emission regions (\object{NGC~4945}, Circinus, etc.) 
from {\it Chandra} images. While there had been indications from the lack of
variability in the soft flux of many AGN that there could be extended X-ray
emission (and a few detections with {\it ROSAT}), the wealth of detail in the
{\it Chandra} images and the correlation with radio and optical properties is
stunning. We now have to understand how this gas is created and the
detailed physics of its relationship to the radio and optical emission
regions (\cite{rmushotzky-C2:ywf01}).

3) The realization that the Fe K lines are often very complex and are 
not adequately modeled by the combination of relativistic effects and
fluorescence in cold material. {\it Chandra} 
data in combination with {\it ASCA} and {\it XMM-Newton} data have 
shown that both narrow lines in combination with broad Fe K
lines are common and that ionized Fe K lines are frequently detected. When
analyzed appropriately (\cite{rmushotzky-C2:yea01a}) the 
{\it Chandra} grating observations
show the existence of complex Fe K line shapes. While it is still early
days it is not yet clear if the narrow Fe K 
components seen by {\it Chandra} are
part of the broad line profile and/or represent emission from a separate
physical region. 
 
4) The proof that Seyfert II galaxies are photoionization dominated --
(\cite{rmushotzky-C2:saea00}) as shown in 
great detail by the {\it Chandra} and {\it XMM-Newton}  grating
observations. However there has not yet been a detailed analysis of the
extended emission lines  revealed by {\it Chandra} images and the spatially
resolved grating data (for a excellent first step see A. Kinkhabwala et al.
this symposium)  

5) The realization that the majority of AGN in the universe do not have
strong broad optical lines as shown by follow-up analysis of the  
{\it XMM-Newton} and {\it Chandra} deep fields 
(e.g., \cite{rmushotzky-C2:bea01}). This, in combination with
serious difference between optical and X-ray classification schemes (based
on {\it Beppo-SAX}, {\it XMM-Newton} and {\it Chandra} serendipitous sources), 
will make a major change in
our understanding of what active galaxies really are, how they evolve and
the distribution of energy with frequency.  

6) The direct X-ray detection of cold material via resonance M shell
absorption in the {\it XMM-Newton} RGS spectra of 
several Seyferts (\cite{rmushotzky-C2:bsk01}).
This was completely unexpected (the first laboratory detection of such an
effect was not until 2001!) and shows the power of X-ray spectroscopy to
simultaneously measure both cold, hot and ionized gas. This unique
capability will become more important in the next few years as more
detailed comparisons between UV, optical and X-ray spectra are
made (\cite{rmushotzky-C2:ck01}).

7) The lack of strong absorption features in Bl Lacs 
from the {\it XMM-Newton} and {\it Chandra} gratings. 
Previous results from {\it Einstein} and low resolution
spectrometers strongly suggested that the X-ray spectra of Bl Lac objects
had strong absorbtion features which so far have not been confirmed. This
extends to the X-ray band the featureless nature of Bl Lac spectra. This
also makes them the best objects to use as white light sources for
measurement of absorption due to the intergalactic  medium. 

8)  The detection of the first features in the power density spectra of active
galaxies with {\it Rossi-XTE} 
and  the lack of simple correlation between UV and X-ray
data (\cite{rmushotzky-C2:en99}; \cite{rmushotzky-C2:poea01}). The estimate of a
characteristic scale in the time domain at a rather low frequency, similar
\
to that seen in galactic black holes will allow a detailed comparison of
these 2 classes of accretion onto black holes. The lack of detailed
correlation in the time domain between the x-ray and the UV is challenging
to most accepted models of active galaxies. 

\section{What observations are needed for progress?}

As seems clear at this meeting, it is very difficult to organize the vast
amount of new information that is being obtained. I believe that we can
make more progress if we consider obtaining and analyzing data along the
following (non-unique) lines. 

1) How do AGN work?
What causes the wide range in Eddington ratios? It is evident that there 
is a very wide range in Eddington ratios for MBHs living in gas rich
environments such as our galactic center or the central regions of
elliptical galaxies. At present there is little or no understanding of this
phenomenon. There may be hope in the detailed study of galactic black hole
transients where the X-ray luminosity varies by a factor of over 100,000.
Also the recent observations (\cite{rmushotzky-C2:kom01}; 
Uttley et al. this symposium)
that individual objects can vary by factors of  over 100 may give some
enlightenment. It is my impression that much of what we think we know about
the X-ray properties of active galaxies have been heavily influenced by
detailed studies of $\sim30$ objects selected as being X-ray bright in the late
1970's. Several of these  seem to have changed their properties
significantly in the last 30 years giving us some insight into the much
larger population of less well studied objects.

Because we now know that most massive galaxies have MBHs one of the
fundamental questions is

a) Why are most MBH quiescent at low z? {\it Chandra} and {\it XMM-Newton} 
observations are able to probe the nuclei of nearby
galaxies down to the limit of a normal X-ray binary and it clear that the
nuclei of some galaxies are even dimmer than this limit. The origins of
this quiescence (most clearly seen in the Milky Way and M31) are not known.
One possibility is that these objects have a ``lifetime'', set by unknown
physics which is much shorter than the age of the universe. 
Given the reasonable estimates of the mass of many nearby MBH from scaling
relations it is clear that there is a very wide range of Eddington ratios
from $L_{Edd}\sim(10^{-7},10^{-5}$) found in low luminosity AGN and LINERs 
and the $L_{Edd}\sim(10^{-3},10^{-1}$) in ``normal'' AGN. In addition to 
the wide range in
Eddington ratios  it is clear that the low luminosity objects have unusual
other properties. For example (\cite{rmushotzky-C2:ptea98}) most low luminosity
objects have less normalized power at high frequencies than normal AGN.
Also the high spatial resolution {\it Chandra} data show that the soft ``thermal''
like emission components seen in many low luminosity AGN originate from
within $1''$ of  the nucleus, distinguishing them from more luminous objects.
These low luminosity objects also tend not to have strong soft continuum components.
Of course, the lowest luminosity of all AGN, \object{NGC~4395}, has rather different
properties than other low luminosity objects (\cite{rmushotzky-C2:moea01}) 

b) Do the X-ray/broad band spectra of these objects match models of low
luminosity objects (such as ADAFs)? 
While there exist excellent sets of time series data for short time scales
we need a better understanding of what the parameters coming out of the
time series analysis mean and better theoretical modeling of the time
series. On long time scales we need much better data with smaller error
bars on  the longer time serie. We also need more well sampled objects, 
good spectra of objects
in ``off'' state, and a good {\it theory of time variability}. The differential
information coming out of the spectral/temporal domain has proven crucial for
understanding galactic black holes and should do the same for AGN. As many
people have noticed AGN are actually brighter per unit time scale than
galactic black holes and so progress is possible.

c) What is the effect of jets  on the broad band spectra and the energetics
of these objects? 
We know that there exist jet dominated sources (Bl Lacs/BLAZARS) and that
emission from jets is  important in several classes of sources (e.g., flat
spectrum radio loud AGN). However the importance of jets in general is
unknown (e.g., \cite{rmushotzky-C2:nwf01}.
We need {\it Chandra} imaging of nearby objects to derive proper samples of
objects with significant X-ray emission from spatially resolved jets.

d) What is the origin of the continuum?
There is no real detailed theory of the origin of the X-ray continuum in
AGN nor do we have guidance for the observational signature of the
fundamental physical mechanism other than the energy dependent time lag
expected in Compton scattering models. While observers have in general fit
simple power laws in the 0.3-50 keV band to the underlying continuum  we do
not really know its true form. As has been clear for many years the X-ray
power law does not extend into the UV band and must steepen at higher
energies. We require broad  band high S/N spectroscopy  with coverage to
higher energies and  analysis of simultaneous UV/xray data. The cutoff at
low frequencies of the X-ray power law is based on the absence of even low
amplitude UV variability in concert with X-ray flares.

e) Is there any direct signature of accretion?
While we have the paradigm that the luminosity is derived from accretion
there is little or no direct evidence for accretion. It is unlikely that
we will find such evidence in the optical/UV band because of the presence
of many very high S/N spectra with so far no direct signature of accretion. 
Is there such a signature in the X-ray spectra? There is a claim of 
highly redshifted absorption in the broad Fe K line in \object{NGC~3516} 
(\cite{rmushotzky-C2:nea99}) which might be a possible
signature of accretion. This must be tested by high S/N spectroscopy Fe K
line spectroscopy, which might only be possible with {\it Astro-E2}. The early
{\it Chandra} and {\it XMM-Newton} spectra do not show any 
evidence for infall in the soft X-ray absorption lines.

2) How are MBHs  connected to their galaxy?  How do they form and evolve?
What are properties of objects with similar mass?

\begin{figure}[ht]
  \begin{center}
    \epsfig{file=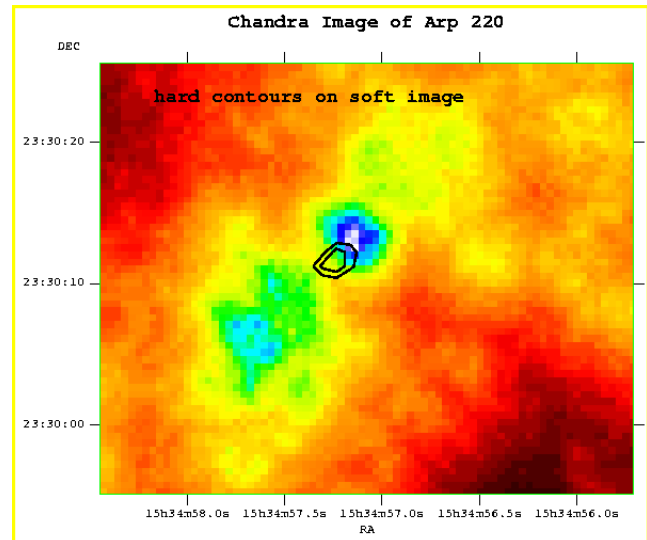, width=8.5cm}
  \end{center}
\caption{The {\it Chandra} soft band image (color) and 
hard band image (contour) of the
famous ultra-luminous infrared selected star forming galaxy Arp 220. The
extended diffuse emission is almost certainly a superwind generated by the 
large amount of star formation. The nature of the low luminosity compact
hard central source is not clear and could be due to either a low luminosity
active nucleus or the sum of X-ray binaries or ultra-luminous sources.}  
\label{rmushotzky-C2_fig:a220}
\end{figure}

In order to pursue these goals we need to isolate the effects of the
Eddington ratio on the X-ray  and broad band spectra and compare the
properties of objects with similar mass at different redshifts and
accretion rates. In order to achieve these goals we need accurate masses of
objects with good X-ray data. This is now possible since we know that the
masses of the central object scale closely with the velocity dispersion of
the stars and that the width of the narrow line region lines scale with the
stellar velocity dispersion (\cite{rmushotzky-C2:nw96}).

a) Why do different types of AGN live in different types of galaxies? 
Is the difference in active galaxy types due to the nature of accretion
which is  from a hot ISM in elliptical galaxies and a cold ISM in spirals ?
Is there any relationship between the nature of the local ISM and that the
central object?

b) What is the star formation/AGN connection? 
Many starburst galaxies have a moderately luminous log $L_X\sim10^{41-42.5}$
ergs s$^{-1}$  hard component, which {\it Chandra} 
images precisely locate in the nucleus (see Figure~\ref{rmushotzky-C2_fig:a220}, 
{\it Chandra} image of \object{Arp 220}). Is this  an AGN, a set of
ultra-luminous X-ray binaries or something else?
There is clearly a relationship between high column densities in the line
of sight and the presence of star formation in many Seyfert IIs. Is the star
formation region the high column density absorber?  In order to settle these 
issues we need high S/N RGS and EPIC {\it XMM-Newton} spectra
and time series of absorbed objects with {\it Chandra} imaging to determine the
nature of the absorber and the emitters.   

c) What is signature of MBH  formation? When and where are the first black
holes forming and what is their spectral signature?  Can deep 
{\it Chandra}/{\it XMM-Newton} 
observations find the ``first black holes'', which should be optically blank
sources in deep fields (\cite{rmushotzky-C2:aea01}). 

3) What is the origin of the wide range of apparent types;
And what causes the difference between them (this is the basis of the
{\it Unified Models})?

a) What are the true Differences between Seyfert Is and IIs? 
An X-ray spectral classification system results in different objects
than an optical one. In particular there are objects that show large X-ray
column densities yet have broad optical lines and there are objects with
narrow optical emission lines that do not show evidence for X-ray
obscuration (\cite{rmushotzky-C2:pea01})  What is the connection and how do the
unified models need to be modified?
We need eigenvector analysis of X-ray/optical/IR samples similar to that
recently done by \cite{rmushotzky-C2:bor02} for the optical data. 

b) What are the X-ray luminous optically ``dull'' 
galaxies (\cite{rmushotzky-C2:eea81}) which make up much of the X-ray background?
Are they MBH without an optical signature or something else? Why are the
optical lines weak/narrow? What is presence of strong broad optical lines
really indicating in optically selected AGN?  What are critical X-ray
observations? 

c) What is the cause of the radio quiet/radio loud dichotomy?  
Why do some have jets?
Is it rapid accretion?, rapidly rotating MBH? Or something else?

The {\it ASCA}/{\it Rossi-XTE} results on radio loud objects show 
systematic differences in
their X-ray spectra, in the frequency of absorption, the strength and shape
of the Fe K  line and in the X-ray spectral slope 
(\cite{rmushotzky-C2:esm00}; \cite{rmushotzky-C2:we87}). It also has long 
been known that radio loud
objects are also systematically X-ray bright. 

We need much better spectral data; larger samples over a range of radio
properties and theoretical models which can provide critical tests of these
free parameters. The mystery of radio loud objects is quite fundamental
since in these objects the bulk of the emitted energy is in the form of
high energy particles rather than radiation and it is only in the X-ray
band that there is a noticeable differences in their radiated
properties.

d) Early results seem to indicate that more massive BHs are more likely to
be radio sources (\cite{rmushotzky-C2:dea01}) -- what is the X-ray connection? 

We need well constructed samples or objects in the radio loudness, BH
mass, X-ray properties space. 

4) How do AGN evolve with cosmic time and flux?
At present we do not know what is actually changing? Is it the accretion
rate, spin, mass, age or perhaps some other parameter.  The {\it ASCA} results
for relatively bright (and therefore luminous objects ) show little
evolution in X-ray spectral properties with z. The early {\it Chandra} results
(\cite{rmushotzky-C2:vea01}) for high redshift objects  indicate that the X-ray
optical ratios for the high z sample  may be smaller by a factor of 2 fainter  
than the low z objects.
However there are subtle issues of bandpass and flux thresholds to take in
to account and so it is not clear if there is a real change. 
If the x-ray to optical ratio is less for the high z objects
it might be that  the earliest quasars are different than lower z objects and if
so we have to search for the origin of the difference. 
The deep surveys show a different redshift  distribution of absorbed vs unabsorbed
sources with the absorbed objects being relatively absent at higher
redshifts. I believe that some of this is a 
 selection effect since the effects of absorption and  reddening becoming
much more severe in the rest frame UV than in the optical band. However it
is clear that the observed redshift distribution of absorbed objects is
lower than that predicted from models of the X-ray background. Absorbed
objects at moderate to high redshift may be almost invisible in the optical
band and indeed this is seen in the progressive reddening of the faint
{\it Chandra} sources.  The general theoretical connection between  star
formation and AGN activity should be most clear in the X-ray band and seems
to have been detected in the {\it Chandra} stacking data for the HDF-N 
(\cite{rmushotzky-C2:nea02}) 

a) {\it XMM-Newton} and {\it Chandra} deep surveys 
show a change in the X-ray luminosity and
optical characteristics of the source population at low fluxes, indicating
that the source counts are dominated by low luminosity spatially resolved
galaxies.  This indicates that  we may have reached the ``end'' of AGN. It
appears that these very faint X-ray objects maybe starforming galaxies as
indicated from the strong correlation with ISO sources. 
(\cite{rmushotzky-C2:aea02}) 
One requires large X-ray samples with good spectra at moderate to high z
combined with optical galaxy properties to determine the nature of these
objects. 

5) What can we learn about strong gravity? The Fe K line shape can be a 
measure of the effects of strong gravity since
much of its flux originates from regions very near the black hole. However
the line shape is  convolved with subtle issues  in modeling the continuum.
This is natural, since the line can be very broad and the bandpass where the
unmodified  ``power law'' continuum can be constrained can be  rather small.
It is not yet clear 
what robust results can be obtained and how to separate out relativistic
effects from modeling uncertainties on the continuum. At the present time
there are only 3 objects with completely unambiguous broad lines; I hope 
that this number will grow rapidly in the near future. 

We do not understand why the line apparently does ``not respond'' to
continuum changes. To quote from a recent study (\cite{rmushotzky-C2:wgy01}) 
{]it ``The general behavior of the line bears little relation to the continuum
...in no source does the Fe K line simply track the continuum''}]. This
statement is true on moderate timescales, but has not, been strongly tested
with CCD data on the characteristic times in which one expects the line to
respond to the continuum, $\sim10$R(G) or $\sim5000$~s in a moderate
luminosity Seyfert galaxy. If indeed the Fe K line does not respond to
continuum changes in either shape and/or flux this will make a fundamental
change in our understanding of the origin of the line. Preliminary detailed
studies of a few objects with {\it XMM-Newton} 
(see Fabian, Reeves, and Pounds, these
proceedings) shows that the Fe K line  does  change in  flux and shape but
the origin of these changes is not known. 
We need high S/N CCD or better time resolved spectra of the Fe K line.
There are only ~10 sources in the sky with a high enough line flux that
even {\it XMM-Newton} can derive the line 
properties on the relevant timescales. We need
long observations of these bright sources to characterize the changes in
the line. Simple estimates show that even for very variable bright sources
$>150$~ks observations are necessary. 

Are there other X-ray spectral features that show relativistic effects?
Given our lack of understanding of the origin of the Fe K lines it is not
clear whether one could or could not predict the existence of other lines with
the signature of  relativistic effects. Thus the possible existence in 
{\it XMM-Newton}
data of broad lines from ionized O, N, and C is exciting 
(\cite{rmushotzky-C2:brea01}. The existence of these lines is at 
present controversial (\cite{rmushotzky-C2:lea01}) 
but in this authors opinion seems likely. It is the combination
of the broad bandpass and high S/N of the 
{\it XMM-Newton} detector combination that made this discovery possible. 
However the lack of broad emission from Fe L, Mg, Si, or S in this 
regard is difficult to understand. We clearly need
the extension of these recent {\it XMM-Newton} RGS results to other 
objects with good signal to noise. 

6) What is the geometry of the central regions?

a) Reflection Component?  Many authors have interpreted the spectral
flattening seen at $E>7$~keV (e.g., \cite{rmushotzky-C2:np94}) as an indication
of Compton reflection from an accretion disk. However the behavior of the
flux of this component with  respect to the driving power law has not been
easily interpreted as originating in an accretion disk. How do we tell if this radiation
is indeed the signature of  a disk? I believe that {\it we need the 
``quality'' of data obtained on galactic black holes} 
(\cite{rmushotzky-C2:gcr01}).
That is {\it sensitive} time resolved spectra of variability over a 
{\it broad} energy band.
While in principle this is possible with a combination of 
{\it XMM-Newton} and {\it Rossi-XTE}/{\it Beppo-SAX}
 so far there are no strong results. Perhaps the  more sensitive
combination of {\it Astro-E2} and {\it INTEGRAL} will be required,

b) What is the warm absorber(WA)? While the WA is an observationally
important component of the soft X-ray spectra we still do not know  if it
is {\it intrinsically interesting or 
just ``in the way''}? By that I mean it is
not yet clear if we  can we learn anything about the nature of the central
object, the nature of accretion, or how the MBH gets fueled from
observations of the WA. So far we have a vast new knowledge of the warm
absorber from {\it Chandra} and {\it XMM-Newton} 
but little ``understanding''. It seems as if
this gas has a low kinetic velocity and a wide range of ionization
indicating that it is low density gas far from the MBH.  It has been
somewhat of a surprise to see that the velocities are less than that of the
broad optical/UV emission lines and similar to that of the UV absorption
lines (\cite{rmushotzky-C2:ck01})

c) Is there a real torus? While schematic models of the central regions
postulate the existence of a molecular torus it is not clear what its
spectral signature is. The  narrow Fe K line seen by 
{\it Chandra} and {\it XMM-Newton} might
be this  spectral signature but it could also be due to gas in the narrow
line or outer broad line region of the AGN or part of the broad line.
Higher spectral resolution with better signal to noise with {\it Astro-E2} and
time variability studies will be required. Other crucial information can be
obtained from the study of sources that have turned off or undergone large
changes in intensity. One might expect that the narrow line will change in
intensity on long time scales potentially allowing a mapping of the region
of origin (\cite{rmushotzky-C2:yea01b}).

d) Can we ever see the accreting material? 
One of the most fundamental postulates of AGN theory is that the emission
is due to accretion, but so far we have no direct evidence. Perhaps we can
detect absorption lines of Fe K line with 
{\it Astro-E2} indicating infalling gas.

\section{Going beyond {\it XMM-Newton} and {\it Chandra}}

In the next few years we will have 3 major new missions: \\
{\it Astro-E2} -- which will have  high sensitivity, high spectral resolution
(E/dE$\sim$700 at Fe K, $\sim160$ at 1 keV), broad band pass (0.3-100 keV)  
and low background. Its relatively poor point spread functions 
is not important for studies of bright or luminous  AGN if 
$F(x)>4\times10^{-14}$ ergs cm$^{-2}$ s$^{-1}$. 
I believe that {\it Astro-E2} will stress detailed Fe K line studies, the
connection of the reflector to Fe K line and spectral time domain studies
with good resolution Its studies of the WA may be limited by resolution \\
{\it INTEGRAL} -- observations will stress the nature of the continuum, 
whether it has a  high energy cutoff, finding ``hard'' AGN  and 
determining the broad band spectra of BLAZARS 

{\it Con-X} -- will provide detailed spectra of wide variety of objects 
and Fe K line reverberation at high S/N spectral resolution. There will be many
10's of objects for which such measurements can be made. The excellent spectral
resolution and high sensitivity will increase the samples of objects with
detailed by spectroscopy by more than a 1000 compared to 
{\it XMM-Newton} and {\it Chandra}. {\it Con-X} will extend x-ray spectroscopy
to the high redshift universe.

\section{Acknowledgements}
I would like to thank T. J. Turner and S. L. Snowden who helped immeasureably with the
manuscript. I would like to thank the XMM project team, the other members of the 
science working group and the inital XMM mission science study group for over 15
years of interesting interactions. Without this large group of committed people 
XMM would still be just an idea. The spectacular results presented at this 
meeting would not have been possible without them. 

\section{Epilogue}

It is interesting to compare what we had hoped from {\it ASCA} with what we
actually achieved. I retrieved a list I made before launch in which I had
listed the important AGN issues: \\
1) Fe line shape measurements of Seyfert galaxies --
We had hoped to establish the existence of a black hole from determining the
detailed Fe line shape and reverberation analysis. 
This was fundamental, new physics and thus part of the original goals for
developing {\it Astro-D}. \\
2) Detailed understanding of Seyfert II galaxies --
Putting the unified model of AGN on a stronger footing  and truly understanding
the nature of Seyfert IIs\\
3) Detailed understanding of the 0.3-2 keV band spectra of Seyfert
galaxies --  What is the nature of the ``soft excess''/WA. 
Is the soft excess a continuum feature (and thus part of the big blue
bump) or are their strong lines due to interesting physics.  Again new
physics was involved  \\
4) Good spectra of ``high z'' QSO's at $E>2$ keV --
Is there evolution in the spectral properties of QSO's, do QSO's differ
from Seyfert I galaxies ? \\
5) Understanding the absorption line spectra of Bl Lac Objects --
What is the absorption due to, is it variable?Obtain  strong bounds on features
due to other elements, try to constrain jet physics (PKS 2155-304)
 \\
6) X-ray Spectra of jets --  Acceleration processes in jets \\
7) Power spectral analysis of variability of Seyfert galaxies at $T<300$~s
-- Search for a characteristic length scale \\
8) Detailed X-ray spectra of ``low'' z qso's \\
9) X-ray spectra of radio galaxies (BLRG, NLRG's , FR type I v II) --  
Nature of unification scheme for Radio loud objects \\
10) the average X-ray spectra of low flux objects and constraints on the
X-ray background


\begin{thebibliography}{}

\bibitem[\protect\astroncite{Alexander et al.}{2001}]{rmushotzky-C2:aea01}
Alexander, D. M., Brandt, W. N., Hornschemeier, A. E., Garmire, G. P., 
Schneider, D. P., Bauer, F. E., Griffiths, R. E. 2001, astro-ph/0107450

\bibitem[\protect\astroncite{Alexander et al.}{2002}]{rmushotzky-C2:aea02}
Alexander, D.M, Aussel, H., Bauer, F. E., Brandt, W. N., Hornschemeier, A. E., 
Vignali, C., Garmire, G. P., Schneider, D. P. 2002, astro-ph/0202493

\bibitem[\protect\astroncite{Antonucci}{1993}]{rmushotzky-C2:ant93}
Antonucci, R. 1993,ARA\&A, 31, 473 

\bibitem[\protect\astroncite{Barger et al.}{2001}]{rmushotzky-C2:bea01}
Barger, A. J., Cowie, L. L., Mushotzky, R. F., Richards, E. A. 2001,
AJ, 121, 662 

\bibitem[\protect\astroncite{Behar, Sako, \& Kahn}{2001}]{rmushotzky-C2:bsk01}
Behar, E., Sako, M., Kahn, S. M. 2001, ApJ, 563, 497 
 
\bibitem[\protect\astroncite{Boroson}{2001}]{rmushotzky-C2:bor02}
Boroson, T. 2002, ApJ, 565, 78

\bibitem[\protect\astroncite{Brandt, Fabian, \& Pounds}{1996}]{rmushotzky-C2:bfp96}
Brandt, W. N., Fabian, A. C., Pounds, K. A. 1996, MNRAS, 278, 326 

\bibitem[\protect\astroncite{Branduardi-Raymont et al.}{2001}]{rmushotzky-C2:brea01}
Branduardi-Raymont, G., Sako, M., Kahn, S. M., Brinkman, A. C., Kaastra, J.
S., Page, M. J 2001, A\&A, 365, 140 

\bibitem[\protect\astroncite{Crenshaw \& Kraemer}{2001}]{rmushotzky-C2:ck01}
Crenshaw, D. M., Kraemer, S. B 2001, ApJ, 562, 29 

\bibitem[\protect\astroncite{Dunlop et al.}{2001}]{rmushotzky-C2:dea01}
Dunlop,  J.S, McLure, R.J., Kukula, M. J., Baum, S. A., O'Dea, C. P., 
Hughes, D. H.  2001, astro-ph/0108397

\bibitem[\protect\astroncite{Edelson \& Nandra}{1999}]{rmushotzky-C2:en99}
Edelson, R., Nandra, K. 1999, ApJ, 514, 682 

\bibitem[\protect\astroncite{Elvis et al.}{1981}]{rmushotzky-C2:eea81}
Elvis, M., Schreier, E. J., Tonry, J., Davis, M., Huchra, J. P.
1981, ApJ, 246, 20 

\bibitem[\protect\astroncite{Eracleous, Sambruna, \& Mushotzky}{2000}]{rmushotzky-C2:esm00}
Eracleous, Michael; Sambruna, Rita; Mushotzky, 
Richard F. 2000, ApJ, 537, 654 

\bibitem[\protect\astroncite{Fabian et al.}{1997}]{rmushotzky-C2:fea97}
Fabian, A. C., Brandt, W. N., McMahon, R. G., Hook, I. M. 1997
MNRAS, 291, 5 

\bibitem[\protect\astroncite{Ferrarese et al.}{2001}]{rmushotzky-C2:fea01}
Ferrarese, L., Pogge, R. W., Peterson, B. M., Merritt, D., Wandel, A., 
Joseph, C. L. 2001, ApJ, 555, 79

\bibitem[\protect\astroncite{Gilfanov, Churazov, \& Revnivtsev}{2001}]{rmushotzky-C2:gcr01}
Gilfanov, M., Churazov, E., Revnivtsev, M. 2001, astro-ph/0001450

\bibitem[\protect\astroncite{Hasinger, et al.}{2001}]{rmushotzky-C2:hea01}
Hasinger, G. et al. 2001, A\&A, 365, 45 

\bibitem[\protect\astroncite{Ho et al.}{2001}]{rmushotzky-C2:hoea01}
Ho, L. C. et al. 2001,ApJ, 549, 51 

\bibitem[\protect\astroncite{Ho, Filippenko, \& Sargent}{1996}]{rmushotzky-C2:hfs96}
Ho, L. C.; Filippenko, A. V., Sargent, Wallace L. W.
1996, ApJ, 462, 183 

\bibitem[\protect\astroncite{Kaastra et al.}{2000}]{rmushotzky-C2:kea00}
Kaastra, J. S., Mewe, R., Liedahl, D. A., Komossa, S., Brinkman, A. C 
2000, A\&A, 354, 83 

\bibitem[\protect\astroncite{Kaspi et al.}{2001}]{rmushotzky-C2:kea01} 
Kaspi, S. et al. 2001, ApJ, 554, 216 

\bibitem[\protect\astroncite{Kennefick et al.}{1997}]{rmushotzky-C2:kea97}
Kennefick, J. D., Osmer, P. S., Hall, P. B., Green, R. F. 
1997, AJ, 114, 2269

\bibitem[\protect\astroncite{Kinkhabwala et al.}{2001}]{rmushotzky-C2:kiea01}
Kinkhabwala, A., Sako, M., Behar, E., Paerels, F., Kahn, S. M. 
2001, tysc conf, 57

\bibitem[\protect\astroncite{Komossa}{2001}]{rmushotzky-C2:kom01}
Komossa, S. 2001,  astro-ph/0109441, ``Lighthouses of the
Universe, the most Luminous Celestial Objects and their use for Cosmology'', 
ESO Astrophysics Symposia, in press

\bibitem[\protect\astroncite{Kormendy \& Gebhardt}{2001}]{rmushotzky-C2:kg01}
Kormendy, J., Gebhardt, K. 2001, astro-ph/0105230, ``The 20$^{th}$
Texas Symposium on Relativistic Astrophysics, ed. H. Martel \& J.C. Wheeler,
AIP, in press

\bibitem[\protect\astroncite{Kotov, Churazov, \& Gilfanov}{2001}]{rmushotzky-C2:kcg01}
Kotov, O., Churazov, E., Gilfanov, M. 2001, MNRAS, 327, 799

\bibitem[\protect\astroncite{La Franca et al.}{2001}]{rmushotzky-C2:lfea01}
La Franca et al. 2001, astro-ph/0112455 2001, ApJ, in press

\bibitem[\protect\astroncite{Levenson, Weaver, \& Heckman}{2001}]{rmushotzky-C2:lwh01}
Levenson, N. A., Weaver, K. A., Heckman, T. M. 2001, ApJ, 550, 230

\bibitem[\protect\astroncite{Lee et al.}{2001}]{rmushotzky-C2:lea01}
Lee, J. C., Ogle, P. M., Canizares, C. R., Marshall, H. L.,
Schulz, N. S., Morales, R., Fabian, A. C., Iwasawa, K.
2001, ApJ, 554, 13 

\bibitem[\protect\astroncite{Loewenstein et al.}{2001}]{rmushotzky-C2:loea01}
Loewenstein, M., Mushotzky, R. F., Angelini, L., Arnaud, K. A.,
Quataert, E. 2001, ApJ, 555, 21

\bibitem[\protect\astroncite{Maiolino et al.}{2000}]{rmushotzky-C2:mea00}
Maiolino, R. et al. 2000, A\&A, 355, 47 

\bibitem[\protect\astroncite{Miller et al.}{2001}]{rmushotzky-C2:mea01}
Miller, J. M. et al. 2001, ApJ, 546, 1055

\bibitem[\protect\astroncite{Miyaji \& Griffiths}{2001}]{rmushotzky-C2:mg01}
Miyaji, T., Griffiths, R. E. 2001, astro-ph/0202048 

\bibitem[\protect\astroncite{Moran et al.}{2001}]{rmushotzky-C2:moea01}
Moran, E. C., Eracleous, M., Leighly, K. M., Chartas, G., Filippenko, A. V.,
Ho, L. C., Blanco, P. R. 2001, astro-ph/0111472, ApJ, Submitted

\bibitem[\protect\astroncite{Mushotzky et al.}{2000}]{rmushotzky-C2:muea00}
Mushotzky, R. F., Cowie, L. L., Barger, A. J., Arnaud, K. A.
2000, Nature, 404, 459

\bibitem[\protect\astroncite{Mushotzky, Done, \& Pounds}{1993}]{rmushotzky-C2:muea93}
Mushotzky, R. F., Done, C., Pounds, K. A. 1993, ARA\&A, 31, 717 

\bibitem[\protect\astroncite{Nagar, Wilson, \& Falcke}{2001}]{rmushotzky-C2:nwf01}
Nagar, N. M., Wilson, A. S., Falcke, H. 2001, ApJ, 559, 87

\bibitem[\protect\astroncite{Nandra \& Pounds}{1994}]{rmushotzky-C2:np94}
Nandra, K., Pounds, K. A. 1994, MNRAS, 268, 405 

\bibitem[\protect\astroncite{Nandra et al.}{1997}]{rmushotzky-C2:nea97}
Nandra, K., George, I. M., Mushotzky, R. F., Turner, T. J., Yaqoob, T.
1997, ApJ, 477, 602

\bibitem[\protect\astroncite{Nandra et al.}{1998}]{rmushotzky-C2:nea98}
Nandra, K., Clavel, J., Edelson, R. A., George, I. M., Malkan, M. A., 
Mushotzky, R. F., Peterson, B. M., Turner, T. J. 1998, ApJ, 505, 594

\bibitem[\protect\astroncite{Nandra et al.}{1999}]{rmushotzky-C2:nea99}
Nandra, K., George, I. M., Mushotzky, R. F., Turner, T. J., Yaqoob, T. 
1999, ApJ, 523, 17

\bibitem[\protect\astroncite{Nandra et al.}{2002}]{rmushotzky-C2:nea02}
Nandra, K., et al. 2002, ??

\bibitem[\protect\astroncite{Nelson \& Whittle}{1996}]{rmushotzky-C2:nw96}
Nelson, C. H., Whittle, M. 1996, ApJ, 465, 96 

\bibitem[\protect\astroncite{Ogle et al.}{2000}]{rmushotzky-C2:oea00}
Ogle, P. M., Marshall, H. L., Lee, J. C., Canizares, C. R.
2000, ApJ, 545, 810

\bibitem[\protect\astroncite{Pappa et al.}{2001}]{rmushotzky-C2:pea01}
Pappa, A., Georgantopoulos, I., Stewart, G. C., Zezas, A. L.
2001, MNRAS, 326, 995 

\bibitem[\protect\astroncite{Petrucci et al.}{2001}]{rmushotzky-C2:peea01}
Petrucci, P. O. et al. 2001, ApJ, 556, 716

\bibitem[\protect\astroncite{Pounds et al.}{2001}]{rmushotzky-C2:poea01}
Pounds, K., Edelson, R., Markowitz, A., Vaughan, S. 2001, ApJ, 550, 15 

\bibitem[\protect\astroncite{Ptak et al.}{1998}]{rmushotzky-C2:ptea98}
Ptak, A., Yaqoob, T., Mushotzky, R., Serlemitsos, P., Griffiths, R. 
1998, ApJ, 501, 37 

\bibitem[\protect\astroncite{Sako et al.}{2000}]{rmushotzky-C2:saea00}
Sako, M., Kahn, S. M., Paerels, F., Liedahl, D. A. 2000, ApJ, 542, 684 

\bibitem[\protect\astroncite{Sambruna et al.}{2001}]{rmushotzky-C2:saea01}
Sambruna, R. M., Maraschi, L., Tavecchio, F., 
Urry, C. M., Cheung, C. C., Chartas, G., Scarpa, R., Reitz, J. K.  
2001, AAS, 1991, 3817 

\bibitem[\protect\astroncite{Smith \& Wilson}{2001}]{rmushotzky-C2:sm01}
Smith, D. A., Wilson, A. S. 2001, ApJ, 557, 180 

\bibitem[\protect\astroncite{Vignali et al.}{2001}]{rmushotzky-C2:vea01}
Vignali, C., Brandt, W. N., Fan, X., Gunn, J. E., Kaspi, S., Schneider, D.
P., Strauss, M. A. 2001, AJ, 122, 2143

\bibitem[\protect\astroncite{Weaver, Gelbord, \& Yaqoob}{2001}]{rmushotzky-C2:wgy01}
Weaver, K. A., Gelbord, J., Yaqoob, T. 2001, ApJ, 550, 261 

\bibitem[\protect\astroncite{Wilkes \& Elvis}{1987}]{rmushotzky-C2:we87}
Wilkes, B. J., Elvis, M. 1987, ApJ, 323, 243 

\bibitem[\protect\astroncite{Wilms et al.}{2001}]{rmushotzky-C2:wea01}
Wilms, J., Reynolds, C. S., Begelman, M. C., Reeves,
J., Molendi, S., Staubert, R., Kendziorra, E. 2001, MNRAS, 328, 27 

\bibitem[\protect\astroncite{Yang, Wilson, \& Ferruit}{2001}]{rmushotzky-C2:ywf01}
Yang, Y., Wilson, A. S., Ferruit, P. 2001, ApJ, 563, 124

\bibitem[\protect\astroncite{Yaqoob et al.}{2001}]{rmushotzky-C2:yea01a} 
Yaqoob, T., George, I. M., Turner, T. J. 2001, astro-ph/0111428, 
in ``High Energy Universe at Sharp Focus: {\it Chandra} Science'',
Eds. E. M. Schlegel and S. Vrtilek, in press

\bibitem[\protect\astroncite{Yaqoob et al.}{2001}]{rmushotzky-C2:yea01b} 
Yaqoob, T., George, I. M., Nandra, K., Turner, T. J., Serlemitsos, P. J.,
Mushotzky, R. F. 2001, ApJ, 546, 759 

\end{thebibliography}
\end{document}